\documentclass{article}
\usepackage{spconf,amsmath,graphicx}
\usepackage{array}
\usepackage{algorithm}
\usepackage{algorithmic}
\usepackage{rotating}
\usepackage{hyperref}
\usepackage{multirow}
\usepackage{booktabs}
\usepackage{longtable}
\usepackage{graphicx}
\usepackage{subfigure}
\title{Learning generic feature representation with synthetic data for weakly-supervised sound event detection by inter-frame distance loss}
\name{Yuxin Huang$^{1,2}$, Liwei Lin$^{1,2}$, Xiangdong Wang$^{1,\dagger}$, Hong Liu$^1$, Yueliang Qian$^1$, Min Liu$^3$, Kazushige Ouchi$^3$}
\address{
  $^1$Beijing Key Laboratory of Mobile Computing and Pervasive Device,\\
  Institute of Computing Technology,
  Chinese Academy of Sciences, Beijing, China\\
  $^2$University of Chinese Academy of Sciences, Beijing, China\\
  \{ huangyuxin18g, linliwei17g, xdwang, hliu, ylqian\}@ict.ac.cn\\
  $^3$Toshiba China R\&D Center, Beijing, China\\
  liumin@toshiba.com.cn, kazushige.ouchi@toshiba.co.jp}
\begin{document}
%\ninept
%
\maketitle
\begin{abstract}

\end{abstract}
Due to the limitation of strong-labeled sound event detection data set, using synthetic data to improve the sound event detection system performance has been a new research focus. In this paper, we try to exploit the usage of synthetic data to improve the feature representation. Based on metric learning, we proposed inter-frame distance loss function for domain adaptation, and prove the effectiveness of it on sound event detection. We also applied multi-task learning with synthetic data. We find the the best performance can be achieved when the two methods being used together. The experiment on DCASE 2018 task 4 test set and DCASE 2019 task 4 synthetic set both show competitive results.

\section{Introduction}
\label{sec:intro}

Sound event detection (SED) is the task to identify the existence and judge the onset and offset of sound events in audio clips. It can be wildly used in applications such as multimedia indexing, security surveillance, autonomous driving and so on \cite{kepler2018large,mcfee2018adaptive}. Due to the high cost of large scale strong labeled training data needed by supervised learning methods for SED \cite{stowell2015detection}, many researchers manage to use only weakly-labeled data, which merely contains the categories of sound events in each audio clip without timestamps of the events \cite{9076321}. It arises the application of weakly-supervised learning in SED, which is usually implemented by multi-instance learning (MIL). 

Another way to solve the problem of lack of annotated data is to use synthetic data. Given a certain number of foreground audio clips of sound events and background audio clips, countless synthetic data with strong labels can be generated by the synthesis procedure \cite{salamon2017scaper}. Because of the advantages of large quantity and strong labels, using synthetic data, especially conjunctively with weakly-labeled data to improve SED systems is becoming a new research focus.

The early approach of using synthetic data is to equate synthetic data with real data. To use synthetic data in weakly-supervised SED, a simple way is to only use the weak labels (category of sound events in an audio clip) of the synthetic data \cite{lin2019guided}, and the strong labels can be further exploited by a sound event detection branch (SEDB) inspired by multi-task learning \cite{huang2020guided}. However, due to the inconsistency between the feature distribution of synthetic data and real data, these approaches are not effective enough. 

To make the feature distribution of synthetic data and real-world data be more close to each other, domain adaptation has been a popular way to use synthetic data in SED \cite{park2019weakly, cornell2020domain}. While using domain adaptation, the model is designed to learn generic characteristic representation which can fit both source domain (synthetic data) and target domain (real-world data), \cite{tzeng2017adversarial} and the extra useful information in source domain can be used in target domain.

Adversary learning is a popular way of domain adaptation. It has been applied in SED for data collecting condition adaptation \cite{wei2020crnn} and jointly learning from weakly-labeled data and synthetic data \cite{park2019weakly, cornell2020domain}. The disadvantage of this approach is that the adversary learning is sensitive to hyperparameters at the training stage, which increases the difficulty of model training.

The other way of domain adaptation is metric learning, the main idea of which is to train the feature encoder to generate similar feature embeddings for similar samples and different feature embeddings for different samples. Many methods has been proposed for metric learning, such as contrastive loss \cite{hadsell2006dimensionality, oord2018representation} and triplet loss \cite{schroff2015facenet}. For these methods, there are three key points: a sampling strategy that decides which samples are used in model training, a distance metric to measure the similarity of feature embeddings and an appropriate model (neural network) structure \cite{qian2015efficient, schroff2015facenet}.  Metric learning has been successfully applied in domain adaptation in fields such image processing \cite{laiz2019using, wang2020triplet}. However, to our knowledge, metric learning is rarely adopted in SED, especially in the exploitation of synthetic data.

In this paper, we explore the usage of metric learning for domain adaptation in SED model training with weakly-labeled and synthetic data, and try to address the three key points mentioned above. A sampling strategy for the polyphonic SED task is proposed, which only uses audio clips with the same sound events and totally different sound events for loss calculation in the metric learning module. To enable frame-level sampling when the real-world data are weakly-labeled, \textit{pseudo strong labels} are generated by fusing the model output with clip-level ground truth. The the distance between frame-level feature embeddings is adopted as distance metric and an inter-frame distance loss is calculated to update the model parameters. For model structure, to perform weakly-supervised learning using the weakly-labeled real-world data, we adopt a MIL framework with a feature encoder by convolutional neural network (CNN) and embedding-level attention pooling. Additionally, an SEDB is also incorporated, which exploits the synthetic data in a multi-task learning way. Experimental results show that the proposed metric learning method using the inter-frame distance loss considerably outperforms the baseline systems that equate the synthetic data with real-world data. It is also shown that incorporating the proposed metric learning approach and the SEDB method can further improve the performance. 

The main contribution of this paper is introduction of metric learning into the application of jointly model training by weakly-labeled and synthetic data. The proposed method solves the problem of choosing sampling strategy and distance metric, and experimental results show the effectiveness of the method as well as the fusion of the method with existing approaches.

\section{Method}
\label{sec:majhead}
\begin{figure}[t]
% \vskip -0.05in
\centering
\includegraphics[width=1.05\linewidth]{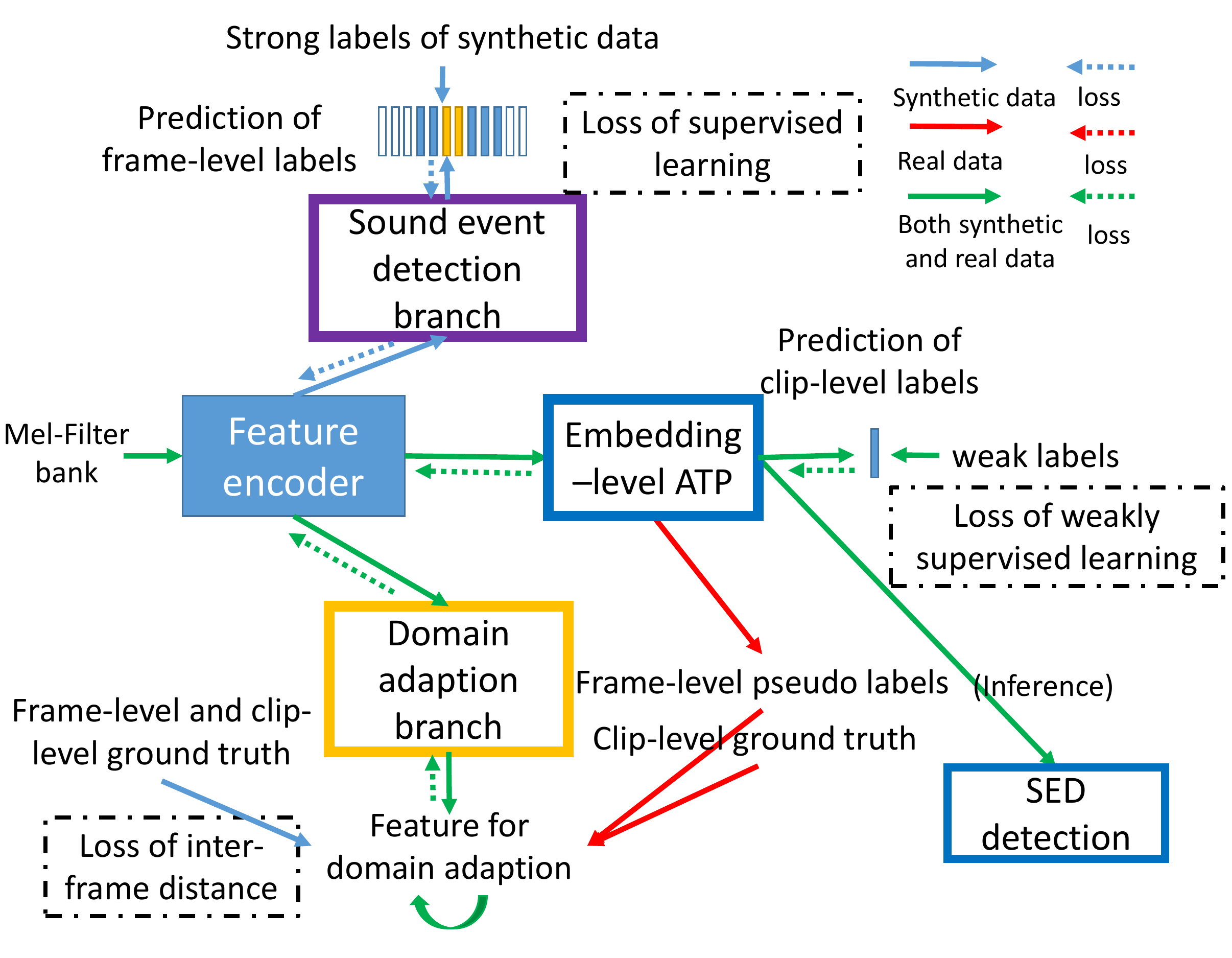}
% \vskip -0.12in
\caption{An overview of the architecture of the model}
\label{fig1}
% \vskip -0.21in
\end{figure}

\begin{figure}
\label{fig_dis}
\begin{minipage}{\linewidth}
  \centering
  \centerline{\includegraphics[width=\linewidth]{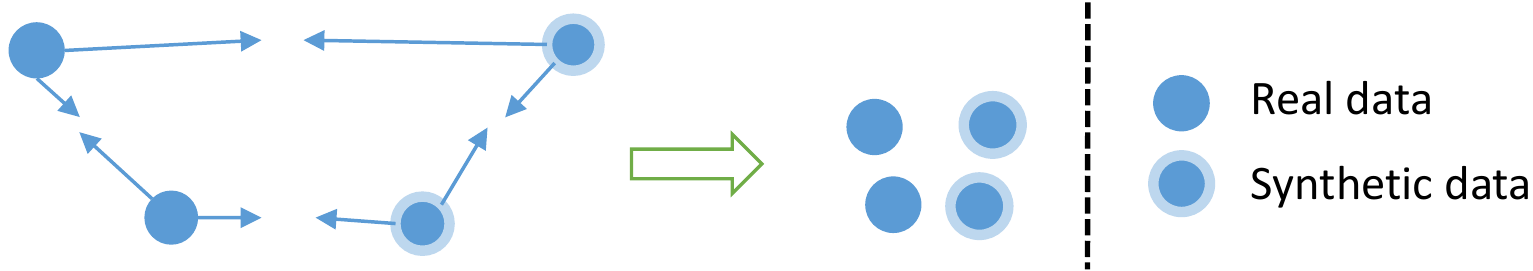}}
  \centerline{(a) Distances variation of positive cases}\medskip
\end{minipage}

\begin{minipage}{\linewidth}
  \centering
  \centerline{\includegraphics[width=\linewidth]{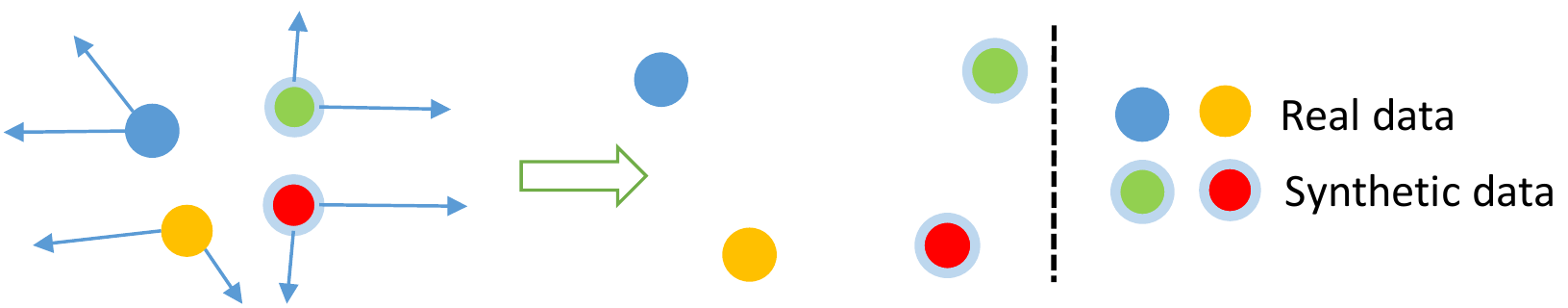}}
  \centerline{(b) Distances variation of negative cases}\medskip
\end{minipage}
\vskip -0.1in
\caption{Distances variation}
\vskip -0.1in
\end{figure}

% \begin{figure}[htbp]
% \centering
% \begin{minipage}[t]{0.48\textwidth}
% \centering
% \includegraphics[width=6cm]{test1.jpg}
% \caption{World Map}
% \end{minipage}
% \begin{minipage}[t]{0.48\textwidth}
% \centering
% \includegraphics[width=6cm]{test2.jpg}
% \caption{Concrete and Constructions}
% \end{minipage}
% \end{figure}

\begin{figure}[htbp]
\centering

\subfigure[Positive cases]{
\begin{minipage}[t]{0.39\linewidth}
\centering
\includegraphics[width=\linewidth]{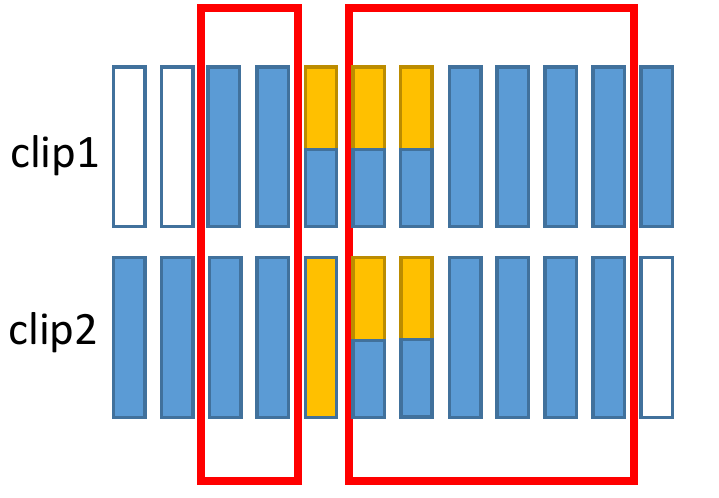}
%\caption{fig1}
\end{minipage}%
}%
\subfigure[Negative cases\  \  \ \ \ \  \ \ \  \ \  \ \ \  \ \ \ \  \   \  \  \   \  \  \ ]{
\begin{minipage}[t]{0.61\linewidth}
\centering
\includegraphics[width=\linewidth]{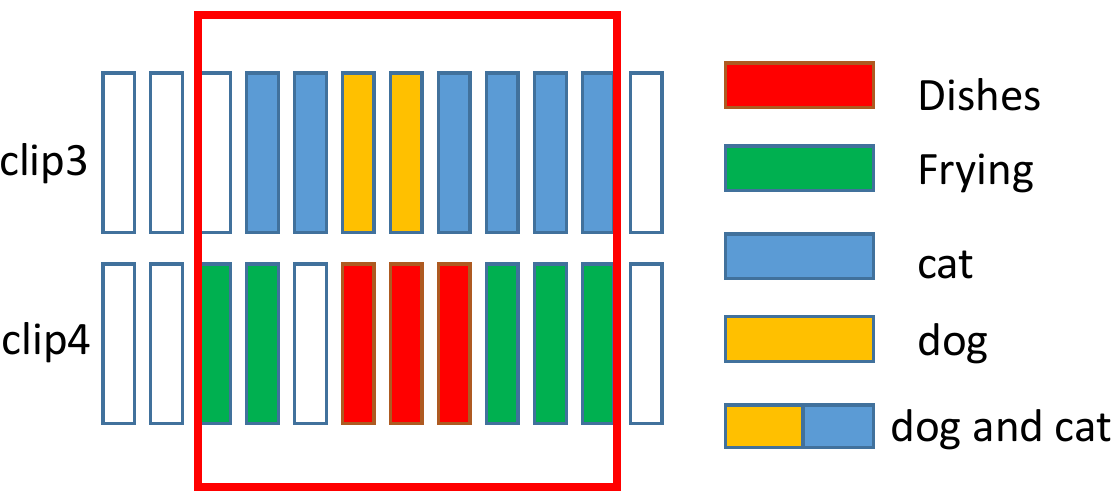}
%\caption{fig2}
\end{minipage}%
}%
\centering
\vskip -0.15in
\caption{ Sampling cases}
\label{fig_cases}
\vskip -0.15in
\end{figure}

% \begin{figure}[htbp]
% \label{fig_cases}
% \begin{minipage}[t]{0.48\textwidth}
%   \centering
%   \includegraphics[width=3cm]{pos_2.pdf}
%   \centerline{(a) Positive cases}
% \end{minipage}

% \begin{minipage}[t]{0.48\textwidth}
%   \centering
%   \includegraphics[width=3cm]{neg_2.pdf}
%   \centerline{(b) Negative cases}
% \end{minipage}

% \vskip -0.1in
% \caption{Sampling cases}
% \vskip -0.1in
% \end{figure}

% \begin{figure}[t]
% % \vskip -0.05in
% \centering
% \includegraphics[width=0.9\linewidth]{pos.pdf}
%  \vskip -0.12in
% \caption{Positive cases}
% \label{fig2}
% \vskip -0.11in
% \end{figure} 

% \begin{figure}[t]
% % \vskip -0.05in
% \centering
% \includegraphics[width=0.9\linewidth]{neg.pdf}
%  \vskip -0.12in
% \caption{Negative cases}
% \label{fig3}
% \vskip -0.15in
% \end{figure} 

    % Our method focus on using synthetic data to get a more generic representation featrue for SED. Two modules are proposed, the domain adaptation branch, which uses the inter-frame distance loss, and the sound event detection branch.

\subsection{Model overview}
The main framework of our model is as shown in Fig.\ref{fig1}. As shown in the legend of data flows at the upper right, in our method, two datasets are used. The first dataset contains real-world data with weak labels, and the second data contains synthetic data with strong labels. At the training stage, both the two kinds of data go through the common feature encoder and are converted into high-level feature embeddings. Then, the feature embeddings are input into three branches: The first branch is the embedding-level attention pooling (ATP) module \cite{9076321}, which performs weakly-supervised learning using both real-world and synthetic data (for the synthetic data, only the weak labels are used in this module). The second branch is the inter-frame domain adaptation module, which is based on the metric learning method. And the third branch is the SEDB, which only uses the synthetic data and performs supervised learning. Only the first branch, namely the embedding-level attention pooling module is used for the test (prediction) process, and the other two branches are used to train the feature encoder.

\subsection{Domain adaptation based on inter-frame distance loss}
Through the common feature encoder and the domain adaptation branch, the real-world data and synthetic data are mapped to the embeddings for domain adaptation. Then the frame-level feature embedding of real-world data and synthetic data are arranged in pairs and domain adaptation based on inter-frame distance (IFD) loss is applied. The domain adaptation is mainly applied to frame-level pairs which are composed of the feature embeddings of a frame in the synthetic data and a frame in the real-world data. It makes the distance of frame-level features of the same category between synthetic data and real-world data be close to each other, and the distance of frame-level features of different categories between synthetic data and real-world data be further. To make the distribution of all data features be more consistent, we also apply the inter-frame domain adaptation to pairs which are composed of only synthetic data or real-world data. An illustration of the distance variation during inter-frame domain adaptation is shown in Fig.\ref{fig_dis}, of which the distances between frame-level feature embeddings with same frame-level label should be closer while the distances between frame-level feature embeddings with different frame-level label should be further. 

The domain adaptation branch is implemented as a dense projection layer, of which the input is the feature embedding output by the feature encoder and the output is the domain embedding representation. The reason for applying domain adaptation branch is to set up a fault tolerant module to avoid the loss of inter-frame domain adaptation over affects the original feature embedding of feature encoder.

As discussed in Section 1, the major challenge of this process is how to choose frame pairs and how to obtain frame-level labels of weakly-labeled data for frame choosing and loss calculation. The following subsections will show our solution in detail.

\subsubsection{Sampling strategy}
\label{ssec:Conditions_of_clustering}
In a batch of data which contains $s$ clips, we composes all clips in pairs and gets $s * s$ pairs for a batch of data. For a single pair, it is represented as $P(c^i_1, c^i_2)$.
Then, in a clip pair, the frames in the two clips are composed as the frame-level pairs. However, in a batch of data, the number of pairs of frame-level feature is large, which costs a lot of computation. To solve this problem, we only sample the frame features at the same time point between two clips as frame-level pairs.

There are three sampling cases:

1) If the clip-level labels of $c^i_1$ and $c^i_2$ are exactly the same, then the distance between the frame features with the same labels in the two clips should be closer, as shown in Fig.\ref{fig_cases} (a).
2) If the clip-level labels of $c^i_1$ and $c^i_2$ are totally different, then the distance between the frame features with different labels in the two clips should be increased, as shown in Fig.\ref{fig_cases} (b).
3) If the audio tags of $c^i_1$ and $c^i_2$ are not exactly the same, the clip-pair is not processed.

\subsubsection{Pseudo label}
\label{ssec:pseudo label}
It can be seen that in order to sample the frame-level feature, it is necessary to have clip-level and frame-level ground truth corresponding to each clip to determine which frame-level features can be clustered and the clustering method. However, in our dataset, only weak labels of the real-world data are available. Therefore, we use the main branch, which is implemented by the embedding-level ATP to generate frame-level pseudo labels for the real-world data domain to implement our method. The frame-level output of the model is fused with the clip-level ground truth to get the final pseudo strong labels for real-world data.
For the synthetic data, the frame-level labels are already provided, so we can use the frame-level labels directly.

\subsubsection{Distance metric and inter-frame distance loss function}
\label{ssec:similarity_measurement}

For measuring distance metric of frames, vector inner product is used as similarity distance. Suppose the vector representation for two frames $v_i$ and $v_j$, the similarity function is:
\begin{equation}
dis_{ij} = \mathbf{v_i}^\mathrm{T}\mathbf{v_j} / \tau(\mathbf{v_i})
\end{equation}
where $\tau(\mathbf{v_i})$ is the number of dimensions of $v_i$. To normalize the distance, we use the loss function for the distance between same frame representation, which is:
\begin{equation}
Loss_{norm} = ||dis_{ii} - 1 ||_2
\end{equation}
We denotes the distances between frames which should be closer (case a in Fig.\ref{fig_cases}) as $dis^{\rm{pos}}$ and the distances between frames which should be further (case b in Fig.\ref{fig_cases}) as $dis^{\rm{neg}}$. So the IFD loss function is:
\begin{equation}
Loss =  \sum_{all\ pairs} {[dis^{\rm{neg}} - dis^{\rm{pos}} + \alpha]_+} + \sum_{all\ pairs}Loss_{norm}
\end{equation}

where$[\cdot]_+ = max(0,\cdot)$, $\alpha$ is the margin, and we set $margin = 0.1$.

\subsection{Sound event detection branch}
\label{ssec:similarity_measurement}
% In previous study, the multi-task learning of SED in which detecting sound event boundaries and deciding the existence of sound events are considered as two tasks are proved to be a good method to improve the performance of SED. However, this method needs data with strong labels for training. In this work, only the synthetic data has strong labels. 

To better exploit the synthetic data with strong labels and increase the diversities of branches, we added a sound event detection branch (SEDB), which had been proposed in \cite{huang2020guided}. The SEDB is inspired by multi-task learning of SED, which considers
detecting sound event boundaries and deciding the existence of sound events as two task. It predicts the frame-level labels of the audio clips , which corresponds to detecting sound event boundaries.
The loss function of SEDB is calculated as:
\begin{equation}
    Loss_{\rm{SEDB}} = \sum_c\sum_t cross\_entropy(y_{ct}, \hat{\mathbf{P}}({y_{ct}}|\mathbf{x_\textit{t}}))
\end{equation}
where $c$ denotes event category, $t$ denotes frame number,  $\hat{\mathbf{P}}({y_{ct}}|\mathbf{x_\textit{t}})$ is the frame-level probability output by the SEDB and $y_{ct}$ is the frame-level ground truth.

\section{Experiment}
\label{sec:print}
\subsection{Dataset}
% For real data, we use the DCASE 2018 Task 4 dataset. For synthetic data, we use the data generated by ourselves and the DCASE 2019 synthetic dataset.
\subsubsection{Real-world data}
The DCASE 2018 Task 4 dataset contains 10 domestic sound event classes.
The dataset is composed of a weakly-labeled training set (1578 clips), an unlabeled training set (54411 clips), a strongly-labeled validation set (288 clips) and a strongly-labeled  test set (880 clips). We take the weakly-labeled set, validation set, test set and do not use the unlabeled set.
\subsubsection{Synthetic data}
The training set of synthetic data is generated with Scaper. The foreground events are obtained from FSD and the background audios are obtained from SINS dataset. All the source audio are provided by DCASE 2020 task 4. The generated synthetic dataset contains 6373 clips. The test set of synthetic data is the synthetic dataset of DCASE 2019 challenge task 4.

% \subsection{Pre-processing and post-processing}
% \label{ssec:pre-post}
% We employ 64 log mel-bank magnitudes extracting from 40 ms frames with 50\% overlap. In this way, all the 10-second audio clips are converted to a feature vector of 500 frames. For post-processing, all the frame-level predictions are smoothed by a adaptive median filter.

\subsection{Training and evaluation}
The feature encoder of the model consists of 3 CNN blocks, which contains a convolution layer, a batch normalization layer and a ReLU activation layer.
We report the event-based marco F1 score \cite{mesaros2016metrics} for evaluating the performance of SED on DCASE 2018 Task 4 test set and DCASE 2019 synthetic dataset. To better analyse the characteristics of each model, we also report the audio tagging F1 score for evaluating the model performance of judging the existence of event.
All the experiments are repeated 20 time with random initialization and we report the average result of each model.

\subsection{Experiment results}

\begin{table}[h]
  \vskip -0.2in
  \caption{The event-based F1 score}
  \vskip 0.1in
  \label{table1}
  \centering
\begin{tabular}{lcc}
\toprule
\textbf{Model} & \textbf{DCASE 2018} & \textbf{DCASE 2019(syn)}\\
\midrule
% The $\mathbf{1^{st}}$ place&-&$0.324$\\
% CDPM \cite{cances2019evaluation}&-& $ 0.320$\\
\hline
baseline &$ 0.315 \pm 0.020 $ &$ 0.320 \pm 0.0191$\\
\hline
IFD &$ 0.331 \pm 0.0218$ &$ 0.361 \pm 0.0162 $\\
\hline
SEDB &$0.336 \pm 0.0169 $ &$ 0.377 \pm 0.0237$\\
\hline
SEDB \& IFD &$\mathbf{0.347 \pm 0.0132 }$ &$\mathbf{0.390 \pm 0.0186 }$ \\
\bottomrule 
\end{tabular}
\vskip -0.42in
\end{table}

\begin{table}[h]
  \caption{The Audio tagging F1 score}
  \vskip 0.1in
  \label{table2}
  \centering
\begin{tabular}{lcc}
\toprule
\textbf{Model} & \textbf{DCASE 2018} & \textbf{DCASE 2019(syn)} \\
\midrule
% The $\mathbf{1^{st}}$ place&-&$0.324$\\
% CDPM \cite{cances2019evaluation}&-& $ 0.320$\\
\hline
baseline &$ 0.625 \pm 0.0141$ & $ 0.630 \pm 0.0328 $\\
\hline
IFD &$ 0.629\pm 0.0199$  &$ 0.609\pm 0.0338 $\\
\hline
SEDB &$0.635 \pm 0.0133 $ & $\mathbf{ 0.663 \pm 0.034}$\\
\hline
SEDB \& IFD &$\mathbf{ 0.645 \pm 0.0118 }$ & $ 0.638 \pm 0.0323 $\\
\bottomrule 
\end{tabular}
\vskip -0.11in
\end{table}

% \begin{table}[h]
%   \caption{The event-based F1 on DCASE 2019 synthetic set }
%   \vskip 0.1in
%   \label{table3}
%   \centering
% \begin{tabular}{lcc}
% \toprule
% \textbf{Model} & \textbf{Average F1 score} \\
% \midrule
% % The $\mathbf{1^{st}}$ place&-&$0.324$\\
% % CDPM \cite{cances2019evaluation}&-& $ 0.320$\\
% \hline
% baseline &$ 0.320 \pm 0.0191 $\\
% \hline
% domain adaptation &$ 0.361 \pm 0.0162 $\\
% \hline
% SEDB &$ 0.377 \pm 0.0237$\\
% \hline
% SEDB \& domain adaptation & $ \mathbf{0.390 \pm 0.0186 }$\\
% \bottomrule 
% \end{tabular}
% \vskip -0.11in
% \end{table}

% \begin{table}[h]
%   \caption{The Audio tagging F1 on DCASE 2019 synthetic set}
%   \vskip 0.1in
%   \label{table4}
%   \centering
% \begin{tabular}{lcc}
% \toprule
% \textbf{Model} & \textbf{Average F1 score}\\
% \midrule
% % The $\mathbf{1^{st}}$ place&-&$0.324$\\
% % CDPM \cite{cances2019evaluation}&-& $ 0.320$\\
% \hline
% baseline & $ 0.630 \pm 0.0328 $\\
% \hline
% domain adaptation &$ 0.609\pm 0.0338 $\\
% \hline
% SEDB & $\mathbf{ 0.663 \pm 0.034}$\\
% \hline
% SEDB \& domain adaptation& $ 0.638 \pm 0.0323 $\\
% \bottomrule 
% \end{tabular}
% \vskip -0.21in
% \end{table}

We conduct experiments on four models. The first model is the baseline model, of which only the embedding-level ATP module is applied. For the second model (denoted as IFD), we add IFD domain adaptation branch to the baseline model. For the third model (denoted as SEDB), we only add SEDB branch to the baseline model. For the fourth model (denoted as SEDB\&IFD), both of the SEDB and domain adaptation module are added.

1) \emph{Can the inter-frame distance domain adaptation improve the model performance?}

For the DCASE 2018 test set and DCASE 2019 synthetic set, we compare both the event-based F1 score and audio tagging F1 score.
As shown in Table\ref{table1}, for DCASE 2018 test set, the baseline system achieves an  event-based F1 score event-based F1 score of 0.315 while the IFD system achieves an event-based F1 score of 0.331.
It can be seen that for event-based F1 score, which is the main evaluation index for SED, the system with domain adaptation can achieves better performance than the baseline system on real-world data set. It proves that the IFD domain adaptation can improve the model performance on real data. In addition, for the event-based F1 score on DCASE 2019 synthetic set, the model with domain adaptation can also achieve better performance than the baseline model, which further prove this conclusion.

To better analyse the model characteristics, we also consider the performance of audio tagging, which evaluate the model performance of judging the existence of event. As shown in Table\ref{table2}, the model with domain adaptation can achieves better performance than the baseline system on DCASE 2018 test set. However, on DCASE 2019 synthetic set, adding domain adaptation brings worse performance than the baseline system. We argue that IFD domain adaptation  focus on make generic frame-level feature representation while the audio tagging is more relied on clip-level feature representation. So, the IFD domain adaptation is more suitable for SED and not for judging the existence of event.

2) \emph{Can the model benefits from both inter-frame domain adaptation and Sound event detection branch?}

We also test the model with both IFD domain adaptation module and SEDB. The model achieves better event-based F1 score than the model with only IFD adaptation module or with SEDB on DCASE 2018 test set and DCASE 2019 synthetic set. We argue that both two model aims to make the feature encoder generate a generic feature representation so that the effectiveness of the two modules can jointly brings a better performance.

\section{Conclusions}
\label{sec:conclusions}

In this paper, we proposed IFD domain adaptation method which is based on metric learning. The experiment results shows it can make generic feature representation for both real data and synthetic data and achieves good result on SED. We also applied SEDB which is based on multi-task learning. The experiment results show that adding both IFD domain adaptation and SEDB achieve greater improvement for SED.

\vfill\pagebreak

% \section{REFERENCES}
% \label{sec:refs}

\bibliographystyle{IEEEbib}
\bibliography{strings,refs}

\end{document}